\documentclass[leqno,12pt]{article}
\usepackage{amsthm}
\usepackage{amssymb}
\usepackage{amsmath}
\usepackage{amsfonts}
\usepackage{enumerate}
\usepackage{verbatim}
\usepackage[dvips]{graphicx}
\usepackage[english]{babel}
\usepackage[cp1250]{inputenc}
\usepackage[T1]{fontenc}
\usepackage[a4paper,left=2.4cm,right=2.4cm,top=2.4cm,bottom=2.4cm]{geometry}
\usepackage{natbib}

\newtheorem{thm}{Theorem}[section]
\newtheorem{lem}{Lemma}[section]

\newtheorem{prop}{Proposition}[section]
\newcommand{\Proof}{\noindent\textbf{{Proof:}}\newline}
\numberwithin{equation}{section}

\newcommand{\cbdu}{\quad\hfill\mbox{$\Box$}\\[3mm]}
\newcommand{\beqo}{\begin{eqnarray*}}
\newcommand{\eeqo}{\end{eqnarray*}\noindent}

\newcommand{\beq}{\begin{eqnarray}}
\newcommand{\eeq}{\end{eqnarray}\noindent}
\def\be{{\mathbb{E}}}
\def\bp{{\mathbb{P}}}
\def\bq{{\mathbb{Q}}}

\def\bfi{{\mathcal{F}}}

\begin{document}

\title{\textbf{Applications of time-delayed backward stochastic differential equations to pricing, hedging and portfolio management}}

\author{\textbf{{\L}ukasz Delong}
\
\\
\footnotesize{Institute of Econometrics, Division of Probabilistic Methods}\\
\footnotesize{Warsaw School of Economics}\\
\footnotesize{Al. Niepodleg{\l}o\'{s}ci 162, 02-554 Warsaw, Poland}\\
\footnotesize{lukasz.delong@sgh.waw.pl}}

\date{}

\maketitle

\newpage

\begin{abstract}
\noindent In this paper we investigate novel applications of a new class of equations which we call time-delayed backward stochastic differential equations. Time-delayed BSDEs may arise in finance when we want to find an investment strategy and an investment portfolio which should replicate a liability or meet a target depending on the applied strategy or the past values of the portfolio. In this setting, a managed investment portfolio serves simultaneously as the underlying security on which the liability/target is contingent and as a replicating portfolio for that liability/target. This is usually the case for capital-protected investments and performance-linked pay-offs. We give examples of pricing, hedging and portfolio management problems (asset-liability management problems) which could be investigated in the framework of time-delayed BSDEs. Our motivation comes from life insurance and we focus on participating contracts and variable annuities. We derive the corresponding time-delayed BSDEs and solve them explicitly or at least provide hints how to solve them numerically. We give a financial interpretation of the theoretical fact that a time-delayed BSDE may not have a solution or may have multiple solutions. 

\end{abstract}

\newpage

\section{Introduction}

\indent Backward stochastic differential equations (BSDEs) have been
introduced in \citet{PP}. Since then, theoretical properties of BSDEs have been thoroughly
studied in the literature and BSDEs have found numerous applications in finance, see for example \citet{K}, \citet{I1}, \citet{pham}.\\
\indent In this paper we study applications of a new class of backward stochastic
differential equation. We consider the dynamics given by
\beq\label{intro}
Y(t)=\xi(Y_T,Z_T)+\int_{t}^{T}f(s,Y_{s},Z_{s})ds-\int_{t}^{T}Z(s)dW(s),\quad 0\leq t\leq T.
\eeq
Here, the generator $f$ at time $s$ and the terminal condition $\xi$ depend on the past
values of a solution $(Y_{s},Z_{s})=(Y(s+u),Z(s+u))_{-T\leq u\leq
0}$. The equation \eqref{intro} can be called a time-delayed backward stochastic differential equation. This type of equations has been introduced and
investigated from the theoretical point of view very recently in \citet{delong1} and \citet{delong3}. Some more theoretical results appear in \cite{gonzalo}. However, no applications have been given so far. The main contribution of this paper is to provide first and novel applications of time-delayed BSDEs to problems related to pricing, hedging and investment portfolio management. We believe that the applications presented in this paper show that time-delayed BSDEs are not only purely theoretical equations but also can be used in applied financial mathematics to solve real-life problems. Additionally, we provide a financial interpretation of the theoretical fact, shown in \citet{delong1}, that a time-delayed BSDE may not have a solution or may have multiple solutions. The theory of time-delayed BSDEs is extended here as well. Compared to \citet{delong1} and \citet{delong3} where the time-delayed values appear in the generator only, we allow the terminal condition in the equation \eqref{intro} to depend on the time-delayed values of a solution as well. Moreover, we succeed in solving some new types of time-delayed BSDEs explicitly and propose a heuristic algorithm to solve time-delayed BSDEs numerically.\\
\indent In financial applications of \eqref{intro}, $Y$ stands for a replicating portfolio, $Z$ denotes a replicating strategy, $\xi$ is a terminal liability, the generator $f$ could model a stream of liabilities which has to be covered over the life-time of a contract. Our main conclusion is that the time-delayed BSDE \eqref{intro} may arise in finance when we want to find an investment strategy $Z$ and an investment portfolio $Y$ which should replicate a liability or meet a target $\xi(Y_T, Z_T), f(s, Y_s, Z_s)$ depending on the applied investment strategy and the past values (past performance) of the investment portfolio. Time-delayed BSDEs may become useful when we face a problem of managing an investment portfolio which serves simultaneously as the underlying security on which the liability/target is contingent and as a replicating portfolio for that liability/target. Non-trivial dependencies of the form $\xi(Y_T, Z_T), f(s,Y_s, Z_s)$ may arise when we consider capital-protected investments and performance-linked pay-offs. We provide specific examples of such pay-offs and financial and insurance problems where pricing, hedging and portfolio management can be studied with the help of a time-delayed BSDE. Notice that the dependence of the value of the liability/target on the replicating portfolio is not present in the classical financial mathematics where the claims are contingent on exogenously given sources of uncertainty and $\xi, f$ do not depend on $(Y,Z)$.\\
\indent Our examples are motivated by the insurance literature. We particularly focus on participating contracts and variable annuities which are life insurance products with capital protections sold worldwide and claims based on the performance of the underlying investment portfolio. In a participating contract the bonuses over a guaranteed return are paid to a policyholder depending on the performance of the insurer's asset portfolio managed by the company's treasurer. In a variable annuity (or a unit-linked product) the insurer faces complicated guarantees providing protection against low or negative returns on the policyholder's investment account which is a collection of different funds. Participating contracts and variable annuities are extensively studied in the actuarial literature but almost all papers treat the insurer's asset portfolio or the policyholder's investment account as an exogenously given stock, beyond the control of the insurer, and assume that bonuses and guarantees are contingent on the performance of that stock, see \citet{bacinello}, \citet{ballotta}, \citet{bauer}, \citet{dai}, \citet{gk}, \citet{milevsky}, \cite{milevskypos} and \citet{milevskyva}. The authors can apply the methods of the classical mathematical finance and consider pricing and hedging of path-dependent European contingent claims.
With their approach dependencies and interactions between the backing investment portfolio and the liability are lost. One eliminates from considerations an important risk management issue as the underlying investment portfolio and its composition can be controlled internally by the insurer who by choosing an appropriate investment strategy can reduce or remove the financial risk of the issued guarantee. \\
\indent The dependence between the applied investment strategy and the pay-off arising under participating contracts has been already noticed in \citet{kleinow2}, \citet{kleinow}, \citet{ballotahab} and \citet{sart}. \citet{kleinow2}, \citet{kleinow} have made a first attempt towards this new kind of investment problem. The authors consider perfect hedging of a participating contract with a guaranteed rate of return and a terminal bonus contingent on the return of a continuously rebalanced asset portfolio which backs the contract liability. However, the derived strategy cannot be financed by the initial premium and the insurer has to provide additional capital to fulfill the obligation arising under the contract. This is clearly a drawback of the approach from \citet{kleinow2}, \citet{kleinow}. In \citet{ballotahab} quadratic hedging is investigated for a participating policy with a terminal benefit equaled to an smoothed value, of an average type, of the assets portfolio held by the insurer. A static investment strategy in the portfolio is found by a numerical experiment. The result is based on a simulation study which, from the mathematical point, is not what we are looking for. Finally, \citet{sart} has constructed for a participating contract an investment portfolio which replicates the benefit contingent on the return earned by that backing portfolio. The investment portfolio consists of bonds being held to maturity and the value of the portfolio is measured at amortized costs. A fixed point problem under which the value of the liability equals the value of the assets held is solved. However, the amortized cost approach, in contrast to our approach under which the rebalancing of assets is possible and the asset value is measured at the current market value, makes the allocation problem in \citet{sart} deterministic. In \citet{sart} the final return on the policy is determined at the inception of the contract and the derived allocation does not allow to create any additional profit by rebalancing the investment portfolio during the term of the policy. This is not a strategy which would be applied in real-life. In this paper we collect the ideas and motivations from \citet{kleinow2}, \citet{kleinow}, \citet{ballotahab}, \citet{sart} and we show how to use time-delayed BSDEs. We remark that in the context of variable annuities or unit-linked products, the optimal composition of assets in the account has not been considered yet.\\
\indent As already mentioned, in financial mathematics most of the problems consider the situation when $\xi, f$ in \eqref{intro} do not depend on $(Y,Z)$. However, capital guarantees and drawdown constraints have attracted the attention of researchers in the context of portfolio management as well. Capital guarantees are practically relevant as protecting the invested capital is the objective of many investors. Under a capital guarantee the value of a managed investment portfolio cannot fall (at the maturity or during the term of the contract) below a prescribed target and under a drawdown constraint the value of a managed investment portfolio cannot fall below a fixed fraction of the running maximum of the past values of the portfolio. In both cases the target for an investment portfolio depends on its past values and the portfolio management problem fits into our proposed framework based on time-delayed BSDEs. We point out the work by \citet{Kjan} where utility maximization and investment strategies under European and American type capital guarantees are investigated. Investment portfolios which fulfill these two types of guarantees are derived and analyzed. An investment problem with a drawdown constraint was solved for the first time in \citet{gz}, and the solution was next extended by \citet{cvitanic}. More recently, \citet{roche} and \citet{elie} consider utility maximization under a drawdown constraint. We further extend the solution from \citet{cvitanic} by constructing an investment portfolio the value of which does not fall, during the time of a contract, below a stochastic fraction of the running maximum of the portfolio accumulating with a predefined growth rate and at the maturity the terminal portfolio value equals the running maximum.\\
\indent Finally, we should recall passport options, see for example \citet{henderson} or \citet{shreve}, and investment problems for a large investor, see for example \citet{buckdahn} and \citet{bank}, to underline the difference of our approach. A passport option also provides a capital protection for an investor. A holder of a passport option (a policyholder) is allowed to change positions in the assets and, regardless of the applied strategy, obtains a non-negative terminal pay-off. In our approach the investment strategy is not chosen by a policyholder but by the financial institution. In the case of a large investor, the applied investment strategy or the wealth has an impact on a stock dynamics and, indirectly, on the liability contingent on that stock. In our examples the investor does not influence a stock dynamics but the liability depends directly on the applied investment strategy.\\
\indent This paper is structured as follows. In Section 2 we introduce time-delayed BSDEs. In Section 3 a motivation and an introduction to pricing, hedging and portfolio management problems in the framework of time-delayed BSDEs are presented with the view towards participating contracts and variable annuities. In Sections 4-7 we deal with ratchet options with gain lock-ins in discrete and continuous time, bonuses and profit sharing based on the average portfolio value and minimum withdrawal rates related to the maximum portfolio value.

\section{Theoretical aspects of time-delayed BSDEs}
\indent We consider a probability space
$(\Omega,\mathcal{F},\mathbb{P})$ with a filtration
$\mathbb{F}=(\mathcal{F}_{t})_{0\leq t\leq T}$ and a finite time
horizon $T<\infty$. We assume that $\mathbb{F}$ is the natural filtration generated by a Brownian motion $W:=(W(t),0\leq t\leq T)$ completed with sets of measure zero. The measure $\mathbb{P}$ is
the real-world, objective probability measure. All equalities or inequalities should be understood in \emph{a.s.} sense.\\
\indent Following \citet{delong1} and \citet{delong3} we introduce a new class of backward equations and we consider the time-delayed backward stochastic equations driven by the Brownian motion of the form
\begin{eqnarray}\label{bsdedelayed}
Y(t)=\xi(Y_T,Z_T)+\int_{t}^{T}f(s,Y_{s},Z_{s})ds-\int_{t}^{T}Z(s)dW(s),\quad
0\leq t\leq T,
\end{eqnarray}
where $\xi$ is the terminal condition of the equation, $f$ is the generator of the equation and the time-delayed values $Y_{s}:=(Y(s+u))_{-T\leq u\leq 0}, Z_{s}:=(Z(s+u))_{-T\leq u\leq 0}$ are fed back into the system.
We always set $Z(t)=0$ and $Y(t)=Y(0)$ for $t<0$. The classical BSDE, without time delays, arises when $\xi$ does not depend on $(Y,Z)$ (is an exogenously given random variable) and $f$ at time $s$ depends only on the value of a solution at this time $f(s,Y(s),Z(s))$ (and an additional exogenously given source of randomness).\\
\indent Let us assume that
\begin{description}
  \item[(A1)] the generator and the terminal condition are product measurable, $\mathbb{F}$-adapted and $\bfi_T$-measurable, and Lipschitz
  continuous, in the sense that for a probability measure $\alpha$ on $[-T,0]\times\mathcal{B}([-T,0])$ and with constants $K_1, K_2$ we have:
  \begin{eqnarray*}
  \lefteqn{\be\Big[|f(\omega,t,Y_{t},Z_{t})-f(\omega,t,\tilde{Y}_{t},\tilde{Z}_{t})|^{2}\Big]}\nonumber\\
  &\leq& K_1 \be\Big[\sup_{0\leq u\leq t}|Y(u)-\tilde{Y}(u)|^{2}+\int_{-T}^{0}|Z(t+u)-\tilde{Z}(t+u)|^{2}\alpha(du)\Big],\\
  \end{eqnarray*}
  and
  \begin{eqnarray*}
  \lefteqn{\be[|\xi(\omega, Y_{T},Z_{T})-\xi(\omega, \tilde{Y}_{T},\tilde{Z}_{T})|^{2}]}\nonumber\\
 &\leq& K_2 \be\Big[\sup_{0\leq u\leq T}|Y(u)-\tilde{Y}(u)|^{2}+\int_{0}^{T}|Z(u)-\tilde{Z}(u)|^{2}\alpha(du)\Big],\\
  \end{eqnarray*}
  for any square integrable processes $(Y,Z)$,$(\tilde{Y},\tilde{Z})$,
  \item[(A2)] $\mathbb{E}\big[\int_{0}^{T}|f(\omega,t,0,0)|^{2}dt\big]<\infty$,
  \item[(A3)] $\be[|\xi(\omega, 0,0)|^2]<\infty$,
  \item[(A4)] $f(\omega,t,.,.)=0$ for $\omega\in\Omega$ and $t<0$.
\end{description}
We emphasize $\omega$ in the definition of $\xi, f$ to remark that the terminal condition and the generator can also depend on some exogenously given randomness driven by $W$. We remark that $f(\omega,t,0,0), \xi(\omega,t,0,0)$ should be understood as the value of the generator and the terminal condition for $Y=Z=0$. We remark that by taking $\alpha$ as Dirac measure we can consider fixed time delays with respect to $Z$, by taking $\alpha$ as Lebesgue measure we can consider delays of an integral form $\int_0^tZ(s)ds$. Notice that we can consider very general delays with respect to $Y$, including fixed delays, delays of an integral form and a maximum over the past values. This generality is due to the introduction of the Lipschitz continuity assumption formulated under the supremum norm.\\
\indent We can prove the following theorem which is an extension of Theorem 2.1 from \citet{delong1}. Compared to \citet{delong1} we assume that a delay appears additionally in the terminal condition and the Lipschitz assumption for $Y$ is formulated under the supremum norm. As the examples of this paper show this modification is useful as it extends possible areas of applications for time-delayed BSDEs.
\begin{thm}\label{thmain}
Assume that \textbf{(A1)}-\textbf{(A4)} hold. For a sufficiently
small time horizon $T$ or for a sufficiently small Lipschitz
constant $K_1$, and for a sufficiently small Lipschitz constant $K_2$, the time-delayed backward stochastic differential equation
\eqref{bsdedelayed} has a unique solution $(Y,Z)$, where the $\mathbb{F}$-adapted process $Y$ and the $\mathbb{F}$-predictable processes $Z$ satisfy
\beq\label{sqint}
\be[\sup_{t\in[0,T]}|Y(t)|^2]<\infty,\quad \be[\int_0^T|Z(s)|^2ds]<\infty.
\eeq
\end{thm}
The proof is similar to the proof which appears in \citet{delong1} and is based on the fixed point theorem in the appropriate Banach space. \\
\indent For the reader's convenience, we recall that under the assumptions of Theorem 2.1, the classical BSDE, without time delays, has a unique solution for arbitrary $T$,$K_1$ and $K_2$, see \citet{K}. We point out that in general we cannot expect to have a unique solution to a time-delayed BSDE for arbitrary parameters $T,K_1, K_2$, see \citet{delong1} for a discussion. This will be clearly illustrated in our examples.

\section{Pricing and hedging with time-delayed BSDEs}
We consider the financial market which consists of two tradeable instruments: a risk-free asset and a risky bond. The price of the risk-free asset $B:=(B(t),0\leq t\leq T)$ is given by the equation
\beq\label{account}
\frac{dB(t)}{B(t)}=r(t)dt,\quad B(0)=1.
\eeq
The risk free interest rate $r:=(r(t),0\leq t\leq T)$ follows a forward It\^{o} diffusion of the form
\beqo
dr(t)=a(t)dt+b(t)dW(t),\quad r(0)=r_0,
\eeqo
where $a:=(a(t),0\leq t\leq T)$,$b:=(b(t),0\leq t\leq T)$ are $\mathbb{F}$-progressively measurable stochastic processes which satisfy usual integrability assumptions. We assume that
\begin{description}
\item[(B1)] the rate $r$ is a non-negative stochastic process with zero as the reflecting barrier.
\end{description}
The price of the risky bond $D:=(D(t),0\leq t\leq T)$ with maturity $T$ is given by
\beq\label{bond}
\frac{dD(t)}{D(t)}=\big(r(t)+\sigma(t)\theta(t)\big)dt+\sigma(t)dW(t),\quad D(0)=d,
\eeq
where $\theta:=(\theta(t),0\leq t\leq T)$,$\sigma:=(\sigma(t),0\leq t\leq T)$ are $\mathbb{F}$-progressively measurable processes which satisfy usual integrability assumptions and under which
\begin{description}
\item[(B2)] $0<D(t)< 1$, $0\leq t< T$, $D(T)=1$,
\end{description}
holds. The process $\theta$ represents the risk premium in the market and we assume that
\begin{description}
\item[(B3)] the process $\theta$ is uniformly bounded a.s. on $[0,T]$.
\end{description}
In particular, there exists a unique equivalent martingale measure $\bq\sim\bp$ under which the discounted price process $D$ is a $(\bq,\mathbb{F})$-martingale. The financial market \eqref{account}-\eqref{bond} is complete and arbitrage-free. It is possible to introduce risky stocks into our model without difficulties. However, as our goal is to illustrate applications of time-delayed BSDEs we decide to keep the model as simple as possible. We denote $\mu(t)=r(t)+\theta(t)\sigma(t)$, $0\leq t\leq T$.\\
\indent Let us consider an investment portfolio $X:=(X(t),0\leq t\leq T)$. Let $\pi:=(\pi(t),0\leq t\leq T)$ denote an amount invested in the bond $D$. Any admissible strategy $\pi$ should be an $\mathbb{F}$-predictable process integrable in It\^{o} sense. The dynamics or the value of the investment portfolio is given by the stochastic differential equation
\beq\label{wealth}
dX(t)=\pi(t)\big(\mu(t) dt+\sigma(t) dW(t))+\big(X(t)-\pi(t)\big)r(t)dt,\quad X(0)=x.
\eeq
By the change of variables
\beq\label{change}
Y(t)=X(t)e^{-\int_0^tr(s)ds},\quad
Z(t)=e^{-\int_0^tr(s)ds}\pi(t)\sigma(t),\quad 0\leq t\leq T\,
\eeq
we arrive at the discounted portfolio process $Y:=(Y(t),0\leq t\leq T)$ under the measure $\bq$
\beq\label{discountedwealth}
dY(t)=Z(t) dW^\bq(t), \quad Y(0)=y,
\eeq
where $W^\bq$ is a $\bq$-Brownian motion. In what follows, we simultaneously work with the undiscounted portfolio $X$ and the discounted portfolio $Y$ and we often recall \eqref{change}.\\
\indent Consider a terminal liability or an investment target $\xi$.  We deal with a problem of finding an investment strategy $\pi$ and an investment portfolio $X$ which replicate a liability or meet a target $\xi(X_T, \pi_T)$ depending on the applied strategy or the past values of the portfolio. In the language of BSDEs our financial problem is equivalent to deriving a solution $(Y,Z)$ to the time-delayed BSDE
\beq\label{wealthbsde}
Y(t)=\tilde{\xi}(Y_T, Z_T)-\int_t^TZ(s)dW^\bq(s),\quad 0\leq t\leq T,
\eeq
which follows immediately from \eqref{discountedwealth}. By \emph{tilde} we always denote discounted pay-offs: $\tilde{\xi}=e^{-\int_0^Tr(s)ds}\xi$.\\
\indent We start by giving a motivating example.\\
\indent \textbf{Example 3.1 - Option Based Portfolio Insurance.} Let us consider an investor who would like to invest $x$ euros. He or she wants to protect the initial capital and hopes to gain an additional profit. Let $S:=(S(t),0\leq t\leq T)$ denote a value process with $S(0)=1$. The process $S$ represents a benchmark or a target investment which the investor would like to follow. In real life $S$ would consist of different traded financial instruments and mutual funds. In the case of our financial market the value process $S$ is contingent on the bank account and bond values $(B, D)$. In the terminology from \citet{Kjan} the process $S$ would be called the value process of an unconstrained allocation. The reader is encouraged to consult \citet{Kjan} for details. It is now well-known that in order to meet the investor's target of protecting the initial capital a financial institution should buy the bond which guarantees $x$ at the terminal time which costs $xD(0)$ and the call option on a fraction $\lambda$ of $S$ which costs $C(x\lambda S(T)-x)=\be^\bq[e^{-\int_0^Tr(s)ds}(x\lambda S(T)-x)^+]$. From \citet{Kjan} we know that there exists a unique $\lambda$ (independent of $x$) such that
\beq\label{structure1}
xD(0)+C(x\lambda S(T)-x)=x,
\eeq
hence the portfolio which fulfills the guarantee of recovering the initial capital $x$ is constructed. By the put-call parity we obtain an equivalent construction under which the amount $x\lambda$ should be invested into the fund $S$ and for the rest of the initial capital we can buy the put option. The fraction $\lambda$ which satisfies \eqref{structure1} also satisfies the following equation
\beq\label{structure2}
x\lambda S(0)+P(x-x\lambda S(T))=x,
\eeq
where $P$ denotes the price of the put option. Both constructions \eqref{structure1} and \eqref{structure2} are called Option Based Portfolio Insurance and gained popularity in financial markets. The fraction $\lambda$ is called a participation factor as the investor participates in the return earned by $S$. In many insurance products, like unit-linked products or variables annuities, the premium which is needed to buy a put option is deducted from the policyholder's account during the life time of a contract, not at the inception as in \eqref{structure2}. With this construction the value of the policyholder's account $V$ is given by the equation
\beqo
dV(t)=V(t)\frac{dS(t)}{S(t)}-V(t)fdt,\quad V(0)=x,
\eeqo
and we easily obtain that $V(t)=xS(t)e^{-ft}$. It is possible to choose a hedging fee $f$ to finance the put option on $(x-V(T))^+$. We obtain that $f=-\ln(\lambda)/T$ where $\lambda$ is defined in \eqref{structure2} and both constructions are again equivalent.\\
\indent The key point is that the above well-known hedging strategy can be obtained by solving a time-delayed BSDE. Let us turn now to dynamic investment strategies based on rebalancing our positions in the bond and the bank account. This might be the only type of a strategy as in many cases static hedging with options cannot be performed due to the lack of such instruments in the market (long durations or not traded underlyings). The goal in our problem is to find $(X,\pi)$ in \eqref{wealth} such that
\beqo
X(T)=X(0)+(X(0)\lambda S(T)-X(0))^+,
\eeqo
which leads to the time-delayed BSDE of the form
\beq\label{example1}
Y(t)&=&e^{-\int_0^Tr(s)ds}\big(Y(0)+(Y(0)\lambda S(T)-Y(0))^+\big)\nonumber\\
&&-\int_t^TZ(s)dW^\bq(s),\quad 0\leq t\leq T,
\eeq
where the terminal condition depends on the past values of $Y$ via $Y(0)$. More generally, we would like to find  $(X,\pi)$ in \eqref{wealth} such that
\beqo
X(T)=X(0)+(X(T)-X(0))^+,
\eeqo
which leads to the time-delayed BSDE of the form
\beq\label{example2}
Y(t)&=&Y(0)e^{-\int_0^Tr(s)ds}+(Y(T)-Y(0)e^{-\int_0^Tr(s)ds})^+\nonumber\\
&&-\int_t^TZ(s)dW^\bq(s),\quad 0\leq t\leq T,
\eeq
where the terminal condition depends on the past values of $Y$ via $Y(0)$ and $Y(T)$. There is a crucial difference between \eqref{example1} and \eqref{example2}. Solving \eqref{example1} is only possible when we know the true composition of $S$ and we are able to replicate $S$ with the bank account and the bond. For the known structure of $S$ and some $\lambda$ we expect to find multiple solutions $(Y,Z)$ to \eqref{example1} which differ in $Y(0)$. The problem \eqref{example1} arises when the target or the liability is set externally. The success of our replication depends on our knowledge about $S$ and availability of the hedging instruments. However, in many financial problems the target or the liability could be contingent on the performance of the internally managed assets and the financial institution can decide on its own $S$. We expect to obtain multiple solutions $(Y, Z)$ to \eqref{example2} which differ on $(Y(0), S)$. The composition for $S$ is dictated by some side criteria relevant to the company offering the product, for example it could be based a utility maximization criterion investigated in \citet{Kjan}.\cbdu
\indent We comment on three possible areas where a time-delayed BSDE \eqref{wealthbsde} may arise.
\indent \textbf{Application 1: Portfolio management.} Example 3.1 clearly shows that some crucial portfolio management problems can be tackled in the framework of time-delayed BSDEs. The key example is a construction of a capital-protected investment as discussed in Example 3.1. This is the first important area of possible applications for our new equations. A simple protection concerns only an initial investment, but a more sophisticated protection could be based on intermediate investment gain lock-ins and portfolio ratcheting or a drawdown constraint. Asian type guarantees paying back an average value of the investment portfolio are also common in portfolio management. Recall that fee schemes in hedge funds are based on hurdle rates and high-water marks which depend on the past values of the underlying investment portfolio which is another area of application for time-delayed BSDEs.\cbdu
\indent \textbf{Application 2: Participating contract.} Participating contracts are modern life insurance endowment contracts which assure maturity payments. The insurer keeps and manages an asset portfolio backing the contractual liability. The asset portfolio is set internally by the treasurer who has the full discretion concerning the allocation and the choice of financial instruments. In our examples there are only two assets available to the treasurer: the bank account and the bond. Under participating policies a policyholder earns a guaranteed return on the initial contribution and participates in the excess return gained by the insurer's asset portfolio. The excess return is usually distributed to a policyholder via a so-called profit sharing scheme. The key feature of participating contracts is that the final pay-off from the policy is related to the performance of the asset portfolio held by the insurer, see TP.2.86-TP.2.93 in \citet{qis}. This implies that the investment strategy applied in the portfolio or the past values of the portfolio have an impact on the final value of the liability. The exceptional feedback between the asset allocation and the terminal benefit makes participating contacts the key example of an insurance product which could be investigated in the framework of time-delayed BSDEs.\cbdu
\indent \textbf{Application 3: Variable annuities and unit-linked products.} Variable annuities and unit-linked contracts are not classified as participating contracts but they share some similarities. Variable annuities and unit-linked products are life insurance investment contracts under which policyholder's contributions are invested in different mutual funds. In our examples there are only two funds available: the bank account and the bond. Positive returns earned on the investment portfolio are distributed to the policyholder's account, but negative returns in the policy are limited due to the fact that these contracts offer guarantees on the policyholder's account value. The key feature of variable annuities and unit-linked products is that the final pay-off is related to the performance of the policyholder's investment account. This again implies that the applied investment strategy in the account or the past values of the investment account have an impact on the final value of the liability. In general, the insurer does not have a discretion concerning the allocation of the capital in the funds, which is up to a policyholder. However, the insurer can propose an investment plan which is very often accepted by a policyholder. In this case the insurer can decide on the allocation in the account. We remark that even if a policyholder can choose its own allocation there are usually some restrictions as the insurer tries to keep some control over the investment accounts.\cbdu
\indent In the cases of participating contracts and variable annuities the goal is to find a composition of the insurer's asset portfolio or the policyholder's investment account under which the guarantee is fulfilled. Moreover, a potential for an additional return should exist under which the profit sharing applies. According to Solvency II Directive, see V.2.2 in \citet{qis}, the insurance reserve must include the estimate of the value of the liability arising under the contract including all possible guarantees and bonuses obtained under the profit sharing scheme. When valuating the liability under a participating contract the future change of the allocations in the backing asset portfolio should be taken into account as the result of future management actions, see TP.2.92 in \citet{qis}. The value of the asset portfolio must match the reserve and the assets held by the insurer must finance the liability which depends on the past and future performance of the asset portfolio and the allocation strategy. A strong relation between the assets and the liability arise as they influence each other. The problems which we investigate in this paper are examples of asset-liability management problems under which the assets and the liabilities must be matched and in our case the condition for matching is of a fixed point nature. We remark that by choosing an investment strategy, applying an appropriate asset-liability strategy, the insurer is able to fulfill the guarantee without deducting any fees needed to buy options and without setting separate hedge accounts as it is considered in the papers mentioned in the Introduction.\\
\indent There are some difficulties related to solving a time-delayed BSDE \eqref{wealthbsde}. There may not be a solution to \eqref{wealthbsde} or there might be a solution which is not interesting from the practical point of view, like a zero or negative solution. These cases are interpreted as impossibility of hedging of a given claim or unfairness of a contract. There might be a unique (practical) solution to \eqref{wealthbsde} which is interpreted as an existence of a unique hedging strategy under a uniquely determined premium (a unique asset-liability strategy). The existence of such solutions could be supported based on Theorem 2.1. However, one has to be careful when applying the fixed point construction as one can end up with a trivial zero solution. Indeed, if $\xi(0,0)=$ then the unique solution arising under Theorem 2.1 is $Y=Z=0$. Finally, there might be multiple solutions to \eqref{wealthbsde}. In many portfolio management problems, including Example 3.1, we would like to find an investment strategy for every initial premium and we should not insist on obtaining a unique solution to a time-delayed BSDE but rather try to recover all multiple solutions. It turns out that in some cases multiple solutions are more meaningful than unique solutions. This is an important difference between time-delayed BSDEs and classical BSDEs. Notice that the existence of multiple solutions in Example 3.1 for \eqref{example1} or \eqref{example2} has a clear financial interpretation as it shows the possibility to meet an investment target or hedge a claim for any initial premium. Interestingly, multiplicity of solutions may also mean an existence of different product designs (different asset-liability strategies) depending on the choice of the value process $S$ as discussed in Example 3.1. We observe all these interpretations in our next examples.\\
\indent In the forthcoming four sections we solve some time-delayed BSDEs. We try to obtain a solution which is square integrable, in the sense of \eqref{sqint}, under the measure $\bq$. Integrability is always understood as integrability under $\bq$ unless stated otherwise. In what follows, we denote participation factors by $\beta> 0, \gamma> 0$ and a guaranteed rate by $g\geq 0$.

\section{Hedging a ratchet contingent on the maximum value of the portfolio - the discrete time case}
In this section we deal with the terminal liability of the form
\beq\label{ratchet1}
\quad \xi=\gamma \max\{X(0)e^{gT},X(t_1)e^{g(T-t_1)}, X(t_2)e^{g(T-t_2)},..., X(t_{n-1})e^{g(T-t_{n-1})},X(T)\},
\eeq
where $g$ is a guaranteed accumulation rate. This is an extension of the capital protection introduced in Example 3.1. The pay-off \eqref{ratchet1} is called a ratchet option and under this protection any intermediate investment gain earned by the investment portfolio $X$ is locked in as the liability and guaranteed to be paid back at maturity. Under \eqref{ratchet1} the highest value of the underlying investment portfolio $X$ over specified anniversaries is paid. Ratchet options are very popular as death or survival benefits in variable annuities. They can also be used as a profit sharing scheme in participating contracts. In portfolio management the target \eqref{ratchet1} would be called a terminal drawdown constraint. We remark that a drawdown constraint appears commonly in portfolio management in the context of capital protecting guarantees.\\
\indent We solve the corresponding time-delayed BSDE.
\begin{prop}
Assume that (B1)-(B3) hold. Consider the time-delayed BSDE of the form
\beq\label{bsderatch1a}
Y(t)&=&\gamma \max\big\{Y(0)e^{-\int_0^Tr(s)ds+gT},Y(t_1)e^{-\int_{t_1}^Tr(s)ds+g(T-t_1)},..., Y(t_{n-1})e^{g(T-t_{n-1})},Y(T)\big\}\nonumber\\
&&-\int_t^TZ(s)dW^\bq(s),\quad 0\leq t\leq T.
\eeq
Let $\mathcal{B}$ and $\mathcal{C}$ denote the following sets
\beqo
\mathcal{B}&=&\Big\{\omega\in\Omega:\gamma\max\big\{e^{gT}D(0),e^{g(T-t_1)}D(t_1),...,e^{g(T-t_{n-1})}D(t_{n-1}),1\big\}>1 \Big\},\\
\mathcal{C}&=&\Big\{\omega\in\Omega:\gamma\max\big\{e^{g(T-t_1)}D(t_1),...,e^{g(T-t_{n-1})}D(t_{n-1}),1\big\}=1 \Big\}.
\eeqo
The equation \eqref{bsderatch1a} has only the following square integrable solutions under the requirement that $Y(0)\geq0$:
\begin{enumerate}
\item[1.] If $\bp(\mathcal{B})>0$ then there exists a unique solution $Y=Z=0$,
\item[2.] If $\bp(\mathcal{B})=0, \gamma e^{gT}D(0)=1$ then there exist multiple solutions $(Y, Z)$, which differ in $Y(0)$, of the form
\beqo
Y(t)&=&\gamma Y(0)e^{gT-\int_0^tr(s)ds}D(t),\quad 0\leq t\leq T,\\
Y(0)e^{gT-\int_0^Tr(s)ds}&=&Y(0)+\int_0^TZ(s)dW(s),
\eeqo
\item[3.] If $\bp(\mathcal{B})=0, \gamma e^{gT}D(0)<1,\bp(\mathcal{C})=1$ then there exist multiple solutions $(Y,Z)$, which differ in $(Y(0),\big(\tilde{\eta}(t_{m+1})\big)_{m=0,1,..n-1})$, of the form
\beqo
Y(t_0)&=&Y(0),\\
Y(t_m)&=&\gamma\max_{k=0,1,...,m}\big\{Y(t_k)e^{-\int_{t_k}^{t_m}r(u)du}e^{g(T-t_k)}\big\}D(t_m)+\be^\bq[\tilde{\eta}(t_{m+1})|\bfi_{t_m}],\\
Y(t_{m+1})&=&\gamma\max_{k=0,1,...,m}\big\{Y(t_k)e^{-\int_{t_k}^{t_{m+1}}r(u)du}e^{g(T-t_{k})}\big\}D(t_{m+1})+\tilde{\eta}(t_{m+1})\\
&&=Y(t_m)+\int_{t_{m}}^{t_{m+1}}Z(s)dW^\bq(s),\\
Y(t)&=&\be^\bq[Y(t_{m+1})|\bfi_t],\quad t_{m}\leq t\leq t_{m+1},\quad m=0,1,...,n-1,
\eeqo
with $t_0=0,t_n=T$ and a sequence of square integrable non-negative random variables $\big(\tilde{\eta}(t_{m+1})\big)_{m=0,1,..n-1}$ such that $\tilde{\eta}(t_{m+1})\in\mathcal{F}_{t_{m+1}}$,
\item[4.] If $\bp(\mathcal{B})=0, \gamma e^{gT}D(0)<1, \bp(\mathcal{C})<1$ then there exists a unique solution $Y=Z=0$.
\end{enumerate}
The solution $Y$ is strictly positive provided that $Y(0)>0$.
\end{prop}
\noindent \textbf{Remark 1:} The requirement $Y(0)\geq 0$ is obvious. It could be omitted but some additional discussion would be needed. Notice that $Y(0)\geq 0$ implies that $Y(t)\geq 0, 0\leq t\leq T$.\\
\noindent \textbf{Remark 2:} The case of $\mathbb{P}(\mathcal{B})=0$ implies that $\gamma\leq 1$ and $\gamma=1$ implies that $\mathbb{P}(\mathcal{C})=1$.\\
\Proof
\noindent 1. As $\mathcal{B}=\bigcup_{m=0,1,...,n}\big\{\omega\in\Omega; \gamma e^{g(T-t_m)}D(t_m)>1\big\}$ there exists $t_k$ with $k=0,1,...,n$ such that $\bp(\gamma e^{g(T-t_k)}D(t_k)>1)>0$. Taking the expected value of \eqref{bsderatch1a} we arrive at the inequality
\beqo
Y(t_k)=\be^\bq[Y(T)|\bfi_{t_k}]\geq \gamma\be^\bq[Y(t_k)e^{-\int_{t_k}^Tr(s)ds+g(T-t_k)}|\bfi_{t_k}]=Y(t_k)\gamma e^{g(T-t_k)}D(t_k),
\eeqo
which results in a contradiction unless $Y(t_k)=0$. As $\be^\bq[Y(T)|\mathcal{F}_{t_k}]=Y(t_k)=0$, by non-negativity of $Y(T)$ we conclude first that $Y(T)=0$ and next that $Y(t)=\be^\bq[Y(T)|\bfi_t]=0,0\leq t\leq T$. Finally, we get that $Z(t)=0$.\\
\noindent 2. We first show that any solution to \eqref{bsderatch1a} must be a $(\bq,\mathbb{F})$-square integrable martingale and fulfill the following representation
\beq\label{estimateeta}
Y(t_{m+1})&=&\gamma\max_{k=0,1,...,m}\big\{Y(t_k)e^{-\int_{t_k}^{t_{m+1}}r(u)du}e^{g(T-t_{k})}\big\}D(t_{m+1})\nonumber\\
&&+\tilde{\eta}(t_{m+1}),\quad m=0,1,...,n-1,
\eeq
with some sequence of non-negative square integrable random variables $(\tilde{\eta}(t_{m+1})_{m=0,1,...,n-1}$ such that $\tilde{\eta}(t_{m+1})\in\mathcal{F}_{t_{m+1}}$. The martingale property of $Y$ is obvious. By taking the expected value in \eqref{bsderatch1a} we arrive at
\beqo
Y(t_{m+1})&=&\gamma\be^\bq\big[\max_{k=0,1,...,n}\big\{Y(t_k)e^{-\int_{t_k}^Tr(u)du}e^{g(T-t_{k})}\big\}|\bfi_{t_{m+1}}\big]\\
&\geq&\gamma\be^\bq\big[\max_{k=0,1,...,m}\big\{Y(t_k)e^{-\int_{t_k}^Tr(u)du}e^{g(T-t_{k})}\big\}|\bfi_{t_{m+1}}\big]\\
&=&\gamma\max_{k=0,1,...,m}\big\{Y(t_k)e^{-\int_{t_k}^{t_{m+1}}r(u)du}e^{g(T-t_{k})}\big\}D(t_{m+1}),
\eeqo
and the statement \eqref{estimateeta} follows. Now we can prove point 2 of our proposition. By taking the expected value of \eqref{estimateeta} we derive
\beq\label{expeta}
Y(0)&=&\be^\bq\big[Y(t_{m+1})\big]\nonumber\\
&=&\be^\bq\big[\gamma\max_{k=0,1,...,m}\big\{Y(t_k)e^{-\int_{t_k}^{t_{m+1}}r(u)du}e^{g(T-t_{k})}\big\}D(t_{m+1})+\tilde{\eta}(t_{m+1})\big],\nonumber\\
&\geq& \gamma Y(0)D(0)e^{gT}+\be^\bq[\tilde{\eta}(t_{m+1})]\nonumber\\
&=&Y(0)+\be^\bq[\tilde{\eta}(t_{m+1})],\quad m=0,1,...,n-1,
\eeq
which immediately implies that $\tilde{\eta}(t_{m+1})=0$ must hold for all $m=0,1,...,n-1$. Assume next that the maximum in \eqref{estimateeta} may not be attained at $t_0=0$. If for some $m=1,...,n-1$
\beqo
\mathbb{Q}\big(\max_{k=0,1,...,m}\big\{Y(t_k)e^{-\int_{t_k}^{t_{m+1}}r(u)du}e^{g(T-t_{k})}\big\}>Y(0)e^{-\int_0^{t_{m+1}}r(u)du}e^{gT}\big)>0,
\eeqo
then we would obtain as in \eqref{expeta} that $Y(0)=\be^\bq\big[Y(t_{m+1})\big]>\gamma Y(0)D(0)e^{gT}$ which results in a contradiction. Hence, $Y(t_{m+1})=\gamma Y(0)e^{gT}e^{-\int_0^{t_{m+1}}r(s)}D(t_{m+1})$ must hold for all $m=0,1,...,n-1$ and a candidate solution $(Y, Z)$ on $[0,T]$ could be defined as in the statement of point 2. It is easy to check that for our candidate solution the terminal condition is fulfilled $\tilde{\xi}=\gamma \max_{m=0,1,...,n}\big\{Y(t_{m})e^{-\int_{t_{m}}^Tr(s)ds+g(T-t_{m})}\big\}=\gamma Y(0)e^{-\int_0^Tr(s)ds+gT}=Y(T)$, hence the solution to \eqref{bsderatch1a} is derived.\\
\noindent 3. The candidate solution follows from the representation \eqref{estimateeta}. From \eqref{estimateeta} we have that $Y(t_1)=\gamma Y(0)e^{-\int_0^{t_1}r(u)du}e^{gT}D(t_1)+\tilde{\eta}(t_1)$ and $Y(0)=\gamma Y(0) e^{gT}D(0)+\be^\bq[\tilde{\eta}(t_1)]$. Hence, we choose $\tilde{\eta}(t_1)$ which ends up with a strictly positive pay-off with a positive probability. Consider $m=1,2,...,n-1$. Following \eqref{expeta} we can derive the relation
\beqo
\lefteqn{\be^\bq[\tilde{\eta}(t_{m+1})|\bfi_{t_m}]}\nonumber\\
&=&Y(t_m)-\gamma\max_{k=0,1,...,m}\big\{Y(t_k)e^{-\int_{t_k}^{t_m}r(u)du}e^{g(T-t_k)}\big\}D(t_m)\nonumber\\
&=&Y(t_m)-\gamma\max\big\{\max_{k=0,1,...,m-1}\big\{Y(t_k)e^{-\int_{t_k}^{t_m}r(u)du}e^{g(T-t_k)}\big\},Y(t_m)e^{g(T-t_m)}\big\}D(t_m)\nonumber\\
&=&Y(t_m)-\max\big\{Y(t_m)-\tilde{\eta}(t_m),\gamma Y(t_m)e^{g(T-t_m)}D(t_m)\big\},\quad m=1,...,n-1,
\eeqo
where we use the representation \eqref{estimateeta}. Now we can conclude that if $\tilde{\eta}(t_m)$ provides a strictly positive pay-off then $\tilde{\eta}(t_{m+1})=0$ if and only if $\gamma e^{g(T-t_m)}D(t_m)=1$, and if $\tilde{\eta}(t_m)$ provides a zero pay-off then $\tilde{\eta}(t_{m+1})=0$. We check whether the constructed solution satisfies the terminal condition. Under the representation of the candidate solution \eqref{estimateeta} we have that
\beqo
Y(T)&=&\gamma\max_{k=0,1,...,n-1}\big\{Y(t_k)e^{-\int_{t_k}^{t_{n}}r(u)du}e^{g(T-t_{k})}\big\}+\tilde{\eta}(t_{n}),
\eeqo
and
\beqo
\tilde{\xi}&=&\gamma\max_{k=0,1,...,n}\big\{Y(t_k)e^{-\int_{t_k}^{t_{n}}r(u)du}e^{g(T-t_{k})}\big\}\\
&=&\gamma\max\big\{\max_{k=0,1,...,n-1}\big\{Y(t_k)e^{-\int_{t_k}^{t_{n}}r(u)du}e^{g(T-t_{k})}\big\},Y(T)\big\}\\
&=&\max\big\{\gamma\max_{k=0,1,...,n-1}\big\{Y(t_k)e^{-\int_{t_k}^{t_{n}}r(u)du}e^{g(T-t_{k})}\big\},\\
&&\gamma\big(\gamma\max_{k=0,1,...,n-1}\big\{Y(t_k)e^{-\int_{t_k}^{t_{n}}r(u)du}e^{g(T-t_{k})}\big\}+\tilde{\eta}(t_{n})\big)\big\}.
\eeqo
Hence $Y(T)=\tilde{\xi}$ if and only if $\gamma=1$ or $\tilde{\eta}(t_{n})=0$. From the earlier considerations it follows that the requirement $\bq(\tilde{\eta}(t_n)=0)=1$ is equivalent to $\bp(\max_{m=1,2,..,n-1}\{\gamma e^{g(T-t_{m})}D(t_m)\}=1)=1$. The candidate solution is the solution to \eqref{bsderatch1a}.\\
\noindent 4. In case 4 we have only zero solution which follows from the discussion concluding the previous point.\\
\noindent Strict positivity of the solution for $Y(0)>0$ is obvious.\cbdu
\indent Hedging of the ratchet on the portfolio \eqref{ratchet1} is only possible if at any time $t$ we end up with the portfolio $X(t)$ which is sufficient to hedge at least its accumulated value $\gamma X(t)e^{g(T-t)}$. If we cannot guarantee that $\gamma e^{g(T-t)}D(t)\leq1$ then the investment portfolio value could fall below the the value of the claim and we could not have enough capital to cover the ratchet (point 1 in Proposition 4.1). A financial institution would not issue the ratchet option under the assumptions of point 1 in the aversion to a shortfall. Hence, zero solution arises. If we set the parameters $(g,\gamma)$ such that the conditions from point 2 in Proposition 4.1 hold then it is possible to hedge the ratchet on the portfolio \eqref{ratchet1} perfectly for any initial premium. Notice that by choosing sufficiently large $g$ and low $\gamma$ the conditions stated in point 2 can always be satisfied. In this case hedging the path-dependent ratchet option on the investment portfolio is equivalent to hedging the fraction $\gamma$ of the guaranteed return $g$ on the initial premium at the terminal time. This investment strategy yields a priori known return related to $(\gamma, g)$ without a potential for an additional profit. This is not a construction which would be implemented in a real life as it is not very appealing to the policyholders. The most important solution to our time-delayed BSDE from the practical point of view is the solution constructed in point 3 in Proposition 4.1. If we set the parameters $(g,\gamma)$ such that the conditions from point hold then it is possible to hedge the ratchet on the portfolio \eqref{ratchet1} perfectly for any initial premium as in point 2. The key point is that in this case there is a potential for an unbounded growth in the portfolio value over the fixed guaranteed return $g$. Notice that by choosing a sufficiently small $g$ the conditions stated in point 3 can always be satisfied. With an appropriate strategy all intermediate investment gains earned by the portfolio once locked-in as the liability can be paid back by liquidating the investment portfolio at the maturity. The specific choice of the sequence $\tilde{\eta}$ is crucial from the practical point of view and could depend on the risk profile of the financial institution and their clients needs. Clearly, the final pay-offs should be attractive to the policyholders. In order to choose $\tilde{\eta}$ the issuer could apply utility maximization, benchmark tracking or mean variance portfolio selection. We do not discuss this step in our paper. The discrete time process $\tilde{\eta}$ is an analogue of the value process of an unconstrained allocation from \citet{Kjan} where the authors suggest to apply utility maximization to determine the target return $\tilde{\eta}$. We refer to \citet{Kjan} for detail discussion and interpretations. We end up with multiple solutions to our equation. Multiplicity of solutions means that the claim can be hedged for any initial premium (point 2) and under different product designs related to different choices of $\tilde{\eta}$ (point 3). Under the assumptions of point 4 in Proposition 4.1 the policyholder could receive only a part of the capital which he or she really owns as $\bp(Y(T)>\tilde{\xi})>0$ holds, see the proof of point 3 in Proposition 4.1. The ratchet option is not fair from the point of view of the policyholder who would not buy it. Again, zero solution arises.\\
\indent We have the following semi-static multi-period Option Based Portfolio Insurance strategy which works under the assumptions of points 2-3 in Proposition 4.1:
\begin{itemize}
\item[1.] At time $t=0$ buy the bond which pays at the terminal time the fraction of the guaranteed return on the available capital. Invest the remaining part of the available capital in an instrument which provides a pay-off $\eta(t_1)\geq 0$ at time $t_1$,
\item[2.] At time $t=t_1$ liquidate your portfolio consisting of the bond providing the fraction of the guaranteed return at $T$ and the instrument paying $\eta(t_1)$. Buy the bond which pays at the terminal time the fraction of the maximum of the guaranteed return on the previously available capital and the currently available capital. Invest the remaining part of the available capital in an instrument which provides a pay-off $\eta(t_2)\geq 0$ at time $t_2$,
\item[3.] Continue till you reach the maturity $T$. Liquidate your portfolio which would cover exactly the ratchet value based on the fraction of the maximum of the past values of the constructed asset portfolio.
\end{itemize}
This is an intuitive extension of the one-period Option Based Portfolio Insurance strategy which has been deduced solely from the solution to the corresponding time-delayed BSDE.

\section{Hedging a ratchet contingent on the maximum value of the portfolio - the continuous time case}
\indent We still consider the liability of a ratchet type contingent on the investment portfolio but now in the continuous time. We investigate the following claim
\beq\label{ratchet2}
\xi=\gamma\sup_{s\in[0,T]}\{X(s)e^{g(T-s)}\}.
\eeq
\indent First, we present the counterparts of points 1-2 from Proposition 4.1.
\begin{prop}
Assume that (B1)-(B3) hold. Consider the time-delayed BSDE of the form
\beq\label{bsderatch1asup}
Y(t)&=&\gamma \sup_{0\leq t\leq T}\{Y(t)e^{-\int_t^Tr(s)ds+g(T-t)}\}\nonumber\\
&&-\int_t^TZ(s)dW^\bq(s),\quad 0\leq t\leq T.
\eeq
Let $\mathcal{D}$ denote the following set
\beqo
\mathcal{D}=\Big\{\omega\in\Omega:\gamma\sup_{0\leq t\leq T}\big\{e^{g(T-t)}D(t)\big\}>1 \Big\}.
\eeqo
The equation \eqref{bsderatch1asup} has only the following square integrable solutions under the requirement that $Y(0)\geq0$:
\begin{enumerate}
\item[1.] If $\bp(\mathcal{D})>0$ then there exists a unique solution $Y=Z=0$,
\item[2.] If $\bp(\mathcal{D})=0, \gamma e^{gT}D(0)=1$ then there exist multiple solutions $(Y, Z)$, which differ in $Y(0)$, of the form
\beqo
Y(t)&=&\gamma Y(0)e^{gT-\int_0^tr(s)ds}D(t),\quad 0\leq t\leq T,\\
Y(0)e^{gT-\int_0^Tr(s)ds}&=&Y(0)+\int_0^TZ(s)dW(s).
\eeqo
\end{enumerate}
The solution $Y$ is strictly positive provided that $Y(0)>0$.
\end{prop}
\noindent \textbf{Remark:} Notice again that the case of $\mathbb{P}(\mathcal{D})=0$ implies that $\gamma\leq 1$.\\
\Proof The proof is analogous to the proof of points 1-2 in Proposition 4.1. We have $\mathcal{D}=\bigcup_{n=1}^{\infty}\bigcup_{m=0}^{n}\{\omega\in\Omega: \gamma e^{g(T-m/nT)}D(m/nT)>1\}$. Under the assumptions of point 1 we deduce that $Y(m/nT)=0$ for some $m,n$ and that $Y=Z=0$ must hold. We can show that any solution $Y$ to \eqref{bsderatch1asup} must fulfill the following representation
\beq\label{etarepresenation2}
Y(t)=\gamma\sup_{0\leq s\leq t}\{Y(s)e^{g(T-s)}e^{-\int_s^tr(u)du}\}D(t)+\tilde{\eta}(t),\quad 0\leq t\leq T,
\eeq
with a square integrable, non-negative and $\mathbb{F}$-adapted process $(\tilde{\eta}(t))_{t\in[0,T]}$. Under the assumptions of point 2 we deduce that $\tilde{\eta}(t)=0$ and $Y(t)=Y(0)e^{gT}e^{-\int_t^Tr(u)du}D(t)$ for $0\leq t\leq T$.\cbdu
\indent In Section 4 we comment that it is always possible to choose $(g,\gamma)$ such that hedging the ratchet on the portfolio \eqref{ratchet1} is equivalent to hedging a fraction of a fixed guaranteed return on the initial premium, see point 2 in Proposition 4.1. In the continuous time we cannot always find $(g,\gamma)$ satisfying the assumptions of point 2 in Proposition 5.1 and hedging of the ratchet on the portfolio \eqref{ratchet2} resulting in the fixed return $\gamma e^{gT}$ may not always be achievable.
\begin{lem}
Let $\mathcal{D}$ be the set defined in Proposition 5.1. In Cox-Ingersoll-Ross interest rate model we cannot find $g\geq0$ and $\gamma= \frac{1}{e^{gT}D(0)}$ such that the condition $\bp(\mathcal{D})=0$ is fulfilled.
\end{lem}
\Proof We have to find $g\geq 0$ to fulfill $\sup_{0\leq t\leq T}\Big\{\frac{e^{-gt}D(t)}{D(0)}\Big\}\leq1$. If it were possible then
\beqo
e^{gt}\geq\frac{e^{n(t)-m(t)r(t)}}{e^{n(0)-m(0)r(0)}},\quad 0\leq t\leq T,
\eeqo
would hold with some continuous functions $n, m$, see Chapter 4.8 in \citet{Cairns}, and equivalently the following condition would hold
\beq\label{cir}
r(t)\geq\frac{gt-n(t)+n(0)-m(0)r(0)}{-m(t)},\quad 0\leq t\leq T.
\eeq
Consider the continuous function $h(t)=\frac{gt-n(t)+n(0)-m(0)r(0)}{-m(t)}$ on $[0,T]$ with $h(0)=r(0)>0$. For any finite $g$, due to continuity of $t\mapsto h(t)$, we have that $h(t)\geq \epsilon>0$ on some small time interval $[0,\lambda]$. However, $r(\lambda)<\epsilon$ with positive probability. Hence, the condition \eqref{cir} is violated with positive probability.\cbdu
\indent The most interesting is an extension of point 3 from Proposition 4.1 which would give us an investment strategy for hedging the ratchet on the portfolio with a potential for an unbounded gain over the guaranteed rate $g$. In the sequel we derive such a strategy. We do not deal with the equation based on the discounted values \eqref{bsderatch1asup} but we return to the equation based on the undiscounted values.\\
\indent First, we show how to construct the process $X$ which satisfies the condition $X(t)\geq \gamma\sup_{s\leq t}\{X(s)e^{g(T-s)}\}D(t)$ which is required to be fulfilled by any solution to the time-delayed BSDE \eqref{bsderatch1asup} as discussed in \eqref{etarepresenation2}. Our next result is an extension of the dynamics under a drawdown constraint from \citet{cvitanic}. Compared to \citet{cvitanic} we require that the controlled process $X$ is above a fraction of its running maximum where the running maximum is based on the process $X$ accumulating with a growth rate not on the process $X$ itself, and the fraction is a stochastic process not a constant.
\begin{lem}
Assume that (B1)-(B3) hold together with
\beqo
\gamma e^{gT}D(0)<1,\quad \gamma\sup_{0\leq t\leq T}\Big\{e^{g(T-t)}D(t)\Big\}\leq 1, \quad \sup_{0\leq t\leq T}|\sigma(t)|\leq K.
\eeqo
Choose an $\mathbb{F}$-predictable process $U$ such that
\beqo
\be\Big[\int_0^T\Big|\frac{U(s)}{D(s)}\Big|^2ds\Big]<\infty,
\eeqo
and consider a non-negative process $S$ under the control $U$ with the forward It\^{o} dynamics
\beqo
dS(t)=U(t)\frac{dD(t)}{D(t)}+(S(t)-U(t))\frac{dB(t)}{B(t)},\quad S(0)=s>0.
\eeqo
There exists a unique square integrable solution under $\bp$ to the forward stochastic differential equation
\beq\label{ratchetcont}
dX(t)&=&\big(\gamma \sup_{0\leq s\leq t}\{X(s)e^{g(T-s)}\}D(t)\big)\frac{dD(t)}{D(t)}\nonumber\\
&&+\big(X(t)-\gamma\sup_{0\leq s\leq t}\{X(s)e^{g(T-s)}\}D(t)\big)\mathbf{1}\{S(t)>0\}\frac{dS(t)}{S(t)},\quad X(0)=x,
\eeq
which fulfills the condition that $X(t)\geq \gamma\sup_{s\leq t}\{X(s)e^{g(T-s)}\}D(t)$ on $[0,T]$.
\end{lem}
\Proof
We follow the idea from \citet{cvitanic}. We deal with the discounted processes $V(t)=\frac{X(t)}{D(t)}$ and $R(t)=\frac{S(t)}{D(t)}$. By It\^{o} formula we obtain the dynamics
\beq\label{dynamicsv}
dV(t)=\big(V(t)-\gamma \sup_{0\leq s\leq t}\{V(s)D(s)e^{g(T-s)}\}\big)\mathbf{1}\{R(t)>0\}\frac{dR(t)}{R(t)},
\eeq
and
\beq\label{dynamicsr}
dR(t)&=&\Big(-R(t)\theta(t)\sigma(t)+R(t)\sigma^2(t)+\frac{U(t)}{D(t)}\theta(t)\sigma(t)-\frac{U(t)}{D(t)}\sigma^2(t)\Big)dt\nonumber\\
&&+\Big(\frac{U(t)}{D(t)}\sigma(t)-R(t)\sigma(t)\Big)dW(t).
\eeq
Let us denote $M(t)=\sup_{0\leq s\leq t}\{\frac{V(s)D(s)e^{g(T-s)}}{D(0)e^{gT}}\}$. Consider the sequence of stopping times $(\tau_n)_{n\in\mathbb{N}}$ defined as $\tau_n=\tau^D_n\wedge\tau^R_n\wedge(T-\frac{1}{n})$ where $\tau^D_n=\inf\{t: \gamma e^{g(T-t)}D(t)=1-\frac{1}{n}\}$ and $\tau^R_n=\inf\{t: R(t)=\frac{1}{n}\}$. We first solve the equation \eqref{ratchetcont} on the time interval $[0,\tau_n]$. We rewrite the dynamics \eqref{dynamicsv} as
\beq\label{eqv}
dV(t)=(V(t)-\gamma M(t)D(0)e^{gT})\frac{dR(t)}{R(t)}
\eeq
By applying It\^{o} formula we can derive
\beqo
d\Big(\frac{V(t)}{M(t)}\Big)&=&\Big(\frac{V(t)}{M(t)}-\gamma D(0)e^{gT}\Big)\frac{dR(t)}{R(t)}-V(t)\frac{dM(t)}{M^2(t)}\\
&=&\Big(\frac{V(t)}{M(t)}-\gamma D(0)e^{gT}\Big)\frac{dR(t)}{R(t)}-\frac{D(0)e^{gt}}{D(t)}\frac{dM(t)}{M(t)},
\eeqo
and
\beqo
d\Big(\log\Big(\frac{V(t)}{M(t)}-\gamma D(0)e^{gT}\Big)\Big)&=&\frac{dR(t)}{R(t)}-\frac{1}{2}\frac{d[R](t)}{R^2(t)}-\frac{1}{\frac{V(t)}{M(t)}-\gamma D(0)e^{gT}}\frac{D(0)e^{gt}}{D(t)}\frac{dM(t)}{M(t)}\\
&=&\frac{dR(t)}{R(t)}-\frac{1}{2}\frac{d[R](t)}{R^2(t)}-\frac{1}{1-\gamma D(t)e^{g(T-t)}}\frac{dM(t)}{M(t)}.
\eeqo
We next obtain the key relation
\beqo
\lefteqn{\log\Big(\frac{V(t)}{M(t)}-\gamma D(0)e^{gT}\Big)-\log\Big(\frac{D(0)}{D(t)}e^{gt}-\gamma D(0)e^{gT}\Big)}\\
&=&\log\big(1-\gamma D(0)e^{gT}\big)-\log\Big(\frac{D(0)}{D(t)}e^{gt}-\gamma D(0)e^{gT}\Big)+\log R(t)-\log R(0)\\
&&-\int_0^t\frac{1}{1-\gamma D(s)e^{g(T-s)}}\frac{dM(s)}{M(s)},\quad 0\leq t\leq \tau_n.
\eeqo
By applying the Skorohod equation we can recover uniquely the processes $(K, L)$ such that
\beq\label{skorohod}
L(t)&=&\log\big(1-\gamma D(0)e^{gT}\big)-\log\Big(\frac{D(0)}{D(t)}e^{gt}-\gamma D(0)e^{gT}\Big)\nonumber\\
&&+\log R(t)-\log R(0),\nonumber\\
K(t)&=&\int_0^t\frac{1}{1-\gamma D(s)e^{g(T-s)}}\frac{dM(s)}{M(s)}=\sup_{0\leq s\leq t}L(t),\nonumber\\
\quad \ L(t)-K(t)&=&\log\Big(\frac{V(t)}{M(t)}-\gamma D(0)e^{gT}\Big)-\log\Big(\frac{D(0)}{D(t)}e^{gt}-\gamma D(0)e^{gT}\Big),
\eeq
hold for $0\leq t\leq \tau_n$. Notice that $L(0)=K(0)=0$ and $K(t)\geq 0$. From the equations \eqref{skorohod} we can derive a unique solution to \eqref{ratchetcont} in the form of
\beq\label{solutiontau}
M(t)&=&V(0)e^{\int_0^t(1-\gamma D(s)e^{g(T-s)})dK(s)},\quad 0\leq t\leq \tau_n,\nonumber\\
V(t)&=&M(t)\Big[\gamma D(0)e^{gT}+(1-\gamma D(0)e^{gT})\frac{R(t)}{R(0)}e^{-K(t)}\Big],\quad 0\leq t\leq \tau_n.
\eeq
Now we would like to extend the solution to $[0,T]$. Consider the processes $(L, K)$ and $(M,V)$ defined in \eqref{skorohod} and \eqref{solutiontau} on the whole interval $[0,T]$. It is straightforward to show that $M(t)\leq V(0)e^{K(t)}$ holds on $[0,T]$ under the condition that $\gamma\sup_{0\leq t\leq T}\{e^{g(T-t)}D(t)\}\leq 1$. Hence we obtain the estimate
\beq\label{psiestimate}
\quad \ 0\leq \psi(t)=V(t)-\gamma M(t)D(0)e^{gT}\leq (1-\gamma D(0)e^{gT})V(0)\frac{R(t)}{R(0)},\quad 0\leq t\leq T.
\eeq
We investigate now the process $\tilde{V}$ defined as
\beqo
\tilde{V}(t)=V(0)+\int_0^{t}\psi(s)\mathbf{1}\{R(s)>0\}\frac{dR(s)}{R(s)},\quad 0\leq t\leq T,
\eeqo
which coincides with the solution \eqref{solutiontau} to the equation \eqref{eqv} on $[0,\tau_n]$. One can show that $\tilde{V}$ is a continuous square integrable semimartingale on $[0,T]$ under the assumptions of our lemma and the derived bound \eqref{psiestimate}.
Hence we obtain the convergence $V(\tau_n)\rightarrow V(\tau_\infty), a.s.$ as $n\rightarrow\infty$ for the process $V$ defined in \eqref{solutiontau}. This also implies the convergence $M(\tau_n)\rightarrow M(\tau_\infty),a.s.$ as $n\rightarrow\infty$ in the view of \eqref{solutiontau}. We can now conclude that the constructed solution $V$ satisfies $V(t)\geq \gamma M(t)D(0)e^{gT}$ on $[0,\tau_\infty]$ which follows from \eqref{solutiontau} again. Notice that if $\tau_\infty<T$ then we must have $R(\tau_\infty)e^{-K(\tau_\infty)}=0$ and we end up with $V(\tau_\infty)=\gamma M(\tau_\infty)D(0)e^{gT}$. This implies that $dV(t)=0$ for $t>\tau_\infty$ and our solution $V$ is defined as constant after $\tau_\infty$. One can easily check that $M(t)=M(\tau_\infty)$ for $t\geq\tau_\infty$. Indeed, we have
\beqo
M(t)&=&\sup_{0\leq s\leq t}\Big\{\frac{V(s)D(s)e^{g(T-s)}}{D(0)e^{gT}}\Big\}=\max\Big\{M(\tau_\infty),\sup_{\tau_\infty\leq s\leq t}\Big\{\frac{V(s)D(s)e^{g(T-s)}}{D(0)e^{gT}}\Big\}\Big\}\\
&=&\max\big\{M(\tau_\infty), M(\tau_\infty)\sup_{\tau_\infty\leq s\leq t}\{\gamma e^{g(T-s)}D(s)\}\big\}=M(\tau_\infty),\quad t\geq\tau_\infty.
\eeqo
The constructed solution $V$ satisfies $V(t)\geq \gamma M(t)D(0)e^{gT}$ on the whole interval $[0,T]$. This gives us that $X(t)\geq \gamma\sup_{s\leq t}\{X(s)e^{g(T-s)}\}D(t)$ holds on $[0,T]$. Finally, square integrability of $X$ is easily deduced from square integrability of $V$.\cbdu
\indent We now give the counterpart of points 3 and 4 from Proposition 4.1
\begin{prop}
Assume that the conditions from Lemma 5.2 hold. Consider the time-delayed BSDE of the form
\beq\label{bsderatch1asup1}
\quad dX(t)&=&\pi(t)\frac{dD(t)}{D(t)}+(X(t)-\pi(t))\frac{dB(t)}{B(t)},\quad X(T)=\gamma\sup_{0\leq s\leq T}\{X(s)e^{g(T-s)}\}.
\eeq
Let $\mathcal{D}$ and $\mathcal{E}$ denote the following sets
\beqo
\mathcal{D}&=&\Big\{\omega\in\Omega:\gamma\sup_{0\leq t\leq T}\big\{e^{g(T-t)}D(t)\big\}>1 \Big\},\\
\mathcal{E}&=&\Big\{\omega\in\Omega:\gamma\sup_{0\leq t\leq T}\big\{e^{g(T-t)}D(t)\big\}=1 \Big\}.
\eeqo
The equation \eqref{bsderatch1asup1} has only the following square integrable solutions under $\bp$ and under the requirement that $X(0)\geq0$:
\begin{enumerate}
\item[3.] If $\bp(\mathcal{D})=0, \gamma e^{gT}D(0)<1, \bp(\mathcal{E})=1$ then there exist multiple solutions $(X,\pi)$, which differ in $(X(0), U)$, of the form
\beqo
X(t)&=&X(0)+\int_0^t\Big(\gamma \sup_{0\leq u\leq s}\{X(u)e^{g(T-u)}\}D(s)\\
&&+\frac{U(s)}{S(s)}\Big(X(s)-\gamma \sup_{0\leq u\leq s}\{X(u)e^{g(T-u)}\}D(s)\Big)\mathbf{1}\{S(s)>0\}\Big)\frac{dD(s)}{D(s)}\\
&&+\int_0^t\Big(1-\frac{U(s)}{S(s)}\Big)\Big(X(s)-\gamma \sup_{0\leq u\leq s}\{X(u)e^{g(T-u)}\}D(s)\Big)\mathbf{1}\{S(s)>0\}\frac{dB(s)}{B(s)},\\
\pi(t)&=&\gamma \sup_{0\leq s\leq t}\{X(s)e^{g(T-s)}\}D(t)\\
&&+\frac{U(t)}{S(t)}\Big(X(t)-\gamma \sup_{0\leq s\leq t}\{X(s)e^{g(T-s)}\}D(t)\Big)\mathbf{1}\{S(t)>0\}
\eeqo
with the process $U$ defined in Lemma 5.2,
\item[4.] If $\bp(\mathcal{D})=0, \gamma e^{gT}D(0)<1, \bp(\mathcal{E})<1$ then there exists a unique solution $Y=Z=0$.
\end{enumerate}
The solution $X$ is strictly positive provided that $X(0)>0$.
\end{prop}
\noindent \textbf{Remark 1:} Notice again that $\gamma=1$ implies that $\mathbb{P}(\mathcal{E})=1$.\\
\noindent \textbf{Remark 2:} If $\sigma$ is not bounded then other assumptions could be formulated to guarantee square integrability of a solution.\\
\Proof
The result follows from Lemma 5.2. We have to investigate the terminal value $V(T)=M(\tau_\infty)\big[\gamma D(0)e^{gT}+(1-\gamma D(0)e^{gT})\frac{R(\tau_\infty)}{R(0)}e^{-K(\tau_\infty)}\big]$ and the terminal condition $\tilde{\xi}=\gamma M(\tau_\infty)D(0)e^{gT}$. It is easy to notice that $V(T)=\tilde{\xi}$ if and only if $R(\tau_\infty)=0$ or $K(\tau_\infty)=+\infty$. The condition $\bp(R(\tau_\infty)=0)=1$ is equivalent to $\bp(\tau^R_\infty\leq T)=1$ which cannot hold as the discounted process $S$ is a $\bq$-martingale which starts at $s>0$. The condition $\bp(K(\tau_\infty)=+\infty)=1$ is equivalent to $\bp(\mathcal{E})=1$.\cbdu
\indent  The process $S$ appearing in Lemma 5.2 and Proposition 5.2 is a continuous time counterpart of the sequence of the random variables $\eta$ from point 3 in Proposition 4.1. The interpretation remains the same. The conclusions for hedging the ratchet are analogous. The conditions stated in point 3 of Proposition 5.2 can be fulfilled by setting a sufficiently small $g$. They are fulfilled by setting $g=0$ trivially. Hence, the ratchet on the portfolio \eqref{ratchet2} can always be hedged if the derived strategy from point 3 is applied in the investment portfolio.

\section{Hedging a pay-off contingent on the average value of the portfolio}
\indent Next to the ratchet studied in the previous sections, a second very common path-dependent pay-off in finance is a pay-off of an Asian type. It is known that a guarantee of locking all intermediate investment gains \eqref{ratchet1}, \eqref{ratchet2} is usually too expensive and sacrifices too much of a potential gain. Instead, to overcome this drawbacks, a pay-off contingent on the average value of the investment portfolio could be introduced. Many participating contracts have profit sharing schemes which are based on an average value of the underlying asset portfolio and such averaging of returns is called smoothing, see \citet{ballotta} and \citet{ballotahab}. There exist variable annuities and unit-linked products in the market under which a bonus as an average value of the policyholder's account is paid additionally at maturity.\\
\indent Let us consider a participating contract or a unit-linked contract which provides a return linked to a benchmark process $S$ and offers a terminal bonus under a profit sharing scheme. The profit sharing scheme is based on the average value of the assets held by the insurer within the duration of the policy. The benchmark process $S$ has the usual interpretation which we discuss in the previous sections and the composition of $S$ is known. We deal with the claim
\beq\label{smoothing}
\xi=X(0)\beta S+\gamma\frac{1}{T}\int_0^Te^{\int_s^Tr(u)du}X(s)ds,
\eeq
where the first term determines the base return related to the terminal value of the benchmark $S:=S(T)$ and the second term represents the bonus triggered by the profit sharing scheme.\\
\indent We derive the corresponding investment strategy.
\begin{prop}
Assume that (B1)-(B3), $S\geq 0, \mathbb{Q}(S>0)>0$ and $\be^\bq[|\tilde{S}|^2]<\infty$ hold. The time-delayed BSDE
\beq\label{smoothbsde}
Y(t)=\beta Y(0) \tilde{S}+\gamma\frac{1}{T}\int_0^TY(s)ds-\int_t^TZ(s)dW^\bq(s),\quad 0\leq t\leq T,
\eeq
has only the following square integrable solutions under the requirement that $Y(0)\geq 0$ and $\int_0^TY(s)ds\geq 0$:
\begin{itemize}
\item[1.] If $\beta\be^{\bq}[\tilde{S}]+\gamma=1$
then there exist multiple solutions $(Y,Z)$, which differ in $Y(0)$, of the form
\beqo
Y(t)=Y(0)+\int_0^tZ(s)dW^\bq(s),\quad 0\leq t\leq T,
\eeqo
with the $\mathbb{F}$-predictable control
\beqo
Z(t)=\frac{1}{1-\gamma+\gamma\frac{t}{T}}M(t),\quad 0\leq t\leq T,
\eeqo
and the process $M$ derived from the martingale representation
\beqo
\beta Y(0) \tilde{S}=\beta Y(0) \be^{\bq}[\tilde{S}]+\int_0^TM(t)dW^\bq(t).
\eeqo
The solution $Y$ is strictly positive provided that $Y(0)>0$.
\item[2.] If $\beta\be^{\bq}[\tilde{S}]+\gamma\neq 1$ then there exists a unique solution $Y=Z=0$.
\end{itemize}
\end{prop}
\noindent \textbf{Remark:} The additional requirement $\int_0^TY(s)ds\geq 0$ is obvious as the bonus paid under the profit sharing scheme cannot be negative. The requirements on $S$ are obvious as well.\\
\Proof
By valuating \eqref{smoothbsde} at $t=0$ we conclude that $(Y,Z)$ must satisfy
\beq\label{smoothbasd2}
Y(0)+\int_0^TZ(s)dW^\bq(s)=\beta Y(0)\tilde{S} +\gamma\frac{1}{T}\int_0^TY(s)ds.
\eeq
By recalling the forward dynamics of the discounted portfolio value \eqref{wealthbsde} we can calculate by Fubini's theorem for stochastic integrals that
\beq\label{average}
\frac{1}{T}\int_0^TY(s)ds&=&\frac{1}{T}\int_0^T\big(Y(0)+\int_0^sZ(u)dW^\bq(u)du\big)ds\nonumber\\
&=&Y(0)+\int_0^T(1-\frac{s}{T})Z(s)dW^{\bq}(s).
\eeq
By substituting the above relation into \eqref{smoothbasd2} we obtain that the pair $(Y,Z)$ must fulfill
\beq\label{smoothbasd3}
Y(0)(1-\gamma)+\int_0^T(1-\gamma+\gamma\frac{s}{T})Z(s)dW^\bq(s)=\beta Y(0) \tilde{S}.
\eeq
\noindent 1. Choose the process $M$ to fulfill the martingale representation of $\beta Y(0)\tilde{S}$ and the process $Z$ to fulfill \eqref{smoothbasd3}. We implicitly assume that $\beta>0$, hence $1-\gamma>0$ holds. The denominator in the definition of $Z$ is strictly positive and square integrability of $M$ implies square integrability of $Z$. We now prove that the constructed solution $Y$ satisfies the requirements of our Proposition. This is trivial if $Y(0)=0$. Assume that $Y(0)>0$. We show that the solution $Y$ is strictly positive. By substituting the derived solution into \eqref{average} we can calculate the average portfolio value as
\beq\label{average1}
\frac{1}{T}\int_0^TY(t)dt&=&Y(0)+\int_0^T\big(1-\frac{t}{T}\big)\frac{1}{1-\gamma+\gamma\frac{t}{T}}M(t)dW^\bq(t)\nonumber\\
&=&Y(0)+\int_0^Th(t)M(t)dW^\bq(t),
\eeq
with
\beqo
h(t)=\frac{T-t}{T-\gamma T+\gamma t}, \quad 0\leq t\leq T.
\eeqo
By the integration by parts formula we obtain that
\beqo
0=h(T)\int_0^TM(t)dW^\bq(t)=\int_0^Th(t)M(t)dW^\bq(t)+\int_0^T\int_0^tM(s)dW^\bq(s)h'(t)dt,
\eeqo
and based on the martingale representation of $\beta Y(0)\tilde{S}$ we can conclude that
\beq\label{averagepositive}
\int_0^Th(t)M(t)dW^\bq(t)&=&-\int_0^T\int_0^tM(s)dW^\bq(s)h'(t)dt\nonumber\\
&=&-\int_0^T(V(t)-V(0))h'(t)dt,
\eeq
where $V(t)=\beta Y(0)\be^\bq[\tilde{S}|\bfi_t]$. By rearranging \eqref{averagepositive} and substituting into \eqref{average1} we arrive at
\beqo
\frac{1}{T}\int_0^TY(s)ds&=&Y(0)+\beta Y(0)\be^\bq[\tilde{S}]\big(h(T)-h(0)\big)-\int_0^TV(t)h'(t)dt\\
&=&-\int_0^TV(t)h'(t)dt> 0.
\eeqo
The inequality is deduced from non-negativity of $V$, continuity of $t\rightarrow V(t)$, strict positivity of $V(0)=\beta Y(0)\be^\bq[\tilde{S}]>0$ and strict negativity of $h'(t)=-\frac{T}{(T+\gamma t-\gamma T)^2}<0$. Strict positivity of $Y$ follows by taking the conditional expected value in \eqref{smoothbsde} together with strict positivity of the participation bonus under the strategy and non-negativity of the benchmark return $\mathbb{E}^\bq[\tilde{S}|\bfi_t]$.\\
\noindent 2. By taking the expected value in \eqref{smoothbasd3} we arrive at a contradiction unless $Y(0)=0$. Taking the expected value in \eqref{smoothbasd2} we obtain that $\mathbb{E}^\bq[\int_0^TY(s)ds]=0$ and by the non-negativity requirement we arrive at $Y(t)=0, 0\leq t\leq T$.\\
\noindent We can allow $\beta=0$ as well and in this case we can conclude that if $\gamma=1$ then $Z(t)=0, Y(t)=Y(0)$, and if $\gamma\neq 1$ then $Z(t)=Y(t)=0$. This is included in our Proposition. \cbdu
\indent We can also provide a similar result for the following claim
\beq\label{smoothings}
\xi=\beta S+\gamma\frac{1}{T}\int_0^Te^{\int_s^Tr(u)du}X(s)ds.
\eeq
\begin{prop}
Assume that (B1)-(B3), $S\geq 0, \mathbb{Q}(S>0)>0$ and $\be^\bq[|\tilde{S}|^2]<\infty$ hold. The time-delayed BSDE
\beq\label{smoothbsde1}
Y(t)=\beta \tilde{S}+\gamma\frac{1}{T}\int_0^TY(s)ds-\int_t^TZ(s)dW^\bq(s),\quad 0\leq t\leq T,
\eeq
has only the following square integrable solutions under the requirement that $Y(0)\geq 0$ and $\int_0^TY(s)ds\geq0$:
\begin{itemize}
\item[1.] If $\gamma<1$
then there exists a unique solution $(Y,Z)$ of the form
\beqo
Y(t)=\frac{\beta\be^\bq[\tilde{S}]}{1-\gamma}+\int_0^tZ(s)dW^\bq(s),\quad 0\leq t\leq T,
\eeqo
with the $\mathbb{F}$-predictable control
\beqo
Z(t)=\frac{1}{1-\gamma+\gamma\frac{t}{T}}M(t),\quad 0\leq t\leq T,
\eeqo
and the process $M$ derived from the martingale representation of
\beqo
\beta \tilde{S}=\beta \be^{\bq}[\tilde{S}]+\int_0^TM(t)dW^\bq(t).
\eeqo
The solution $Y$ is strictly positive provided that $Y(0)>0$.
\item[2.] If $\gamma\geq1$ then there exists no solution.
\end{itemize}
\end{prop}
\noindent \textbf{Remark:} We implicitly assume that $\beta>0$. The case of $\beta=0$ is included in Proposition 6.1.\\
\indent Notice that the investment strategy which we have derived in Proposition 6.1 and which replicates the claim \eqref{smoothing} is to split the initial contribution $X(0)$ into two parts: the first part $\beta X(0)\be^\bq[\tilde{S}]$ is used to buy the benchmark which provides the base return $\beta X(0)S$ at maturity, the second part $\gamma X(0)$ is used to hedge the claim, the participation bonus, arising under the profit sharing scheme. In particular, the claim under the profit sharing scheme includes the Asian guarantee on the benchmark $S$. The investment strategy $U$ for hedging the participation bonus and the value of the corresponding replicating portfolio $G$ can be derived from solving the time-delayed BSDE
\beqo
G(t)=\gamma\beta X(0)\frac{1}{T}\int_0^T\tilde{S}(t)dt+\gamma\frac{1}{T}\int_0^TG(s)ds-\int_t^TZ(s)dW^\bq(s),\quad 0\leq t\leq T,
\eeqo
which is of the form \eqref{smoothbsde1}. The investment portfolio $X$ backing our participating contract is of the form $X(t)=\beta X(0)S(t)+G(t)e^{\int_0^tr(s)ds}$ where $S(t)$ is the value of the benchmark investment providing the base return and $G(t)e^{\int_0^tr(s)ds}$ is the value of the replicating portfolio hedging the participation bonus. Such decomposition is important as Solvency II Directive requires a separate disclosure of the value of all guarantees and participation benefits, see TP.2.87 in \citet{qis}. By the construction, the assets and the liabilities are matched.\\
\indent We finish this section by giving a real-life example of a profit sharing scheme which occurs commonly in participating contract in the UK and which is based on the average return of the underlying asset portfolio. The claim is of the form
\beq\label{ballottaclaim}
\xi&=&X(0)\Big(1+\max\Big\{g,\beta\Big(\frac{X(1)}{X(0)}-1\Big)\Big\}\Big)\cdot\Big(1+\max\Big\{g,\frac{\beta}{2}\Big(\frac{X(1)}{X(0)}+\frac{X(2)}{X(1)}-2\Big)\Big\}\Big)\nonumber\\
&&\cdot\Big(1+\max\Big\{g,\frac{\beta}{T}\Big(\frac{X(1)}{X(0)}+\frac{X(2)}{X(1)}+...+\frac{X(T)}{X(T-1)}-T\Big)\Big\}\Big),
\eeq
and is investigated in \citet{ballotta}, \citet{ballotahab}. In order to find a replicating strategy for \eqref{ballottaclaim} a time-delayed BSDE could be applied. It seems to be a challenging task to find a solution to the corresponding equation which could provide a pay-off strictly above the guaranteed return $g$. This example shows possible further applications of time-delayed BSDEs.

\section{Hedging a stream of payments based on the maximum value of the portfolio}
\indent In this final section we give an example of a claim which leads to an equation with a time delay entering the generator of a BSDE. We investigate a minimum withdrawal benefit. A minimum withdrawal benefit is an important example of a variable annuity contract which is gaining popularity in the market.\\
\indent Under minimum withdrawal schemes the policyholder is allowed to withdraw guaranteed amounts over duration of a contract and receives the remaining value of the account at maturity. Recently, a variable annuity with a minimum withdrawal scheme has been investigated in \citet{milevsky} with a guaranteed withdrawal amount set as a fraction of the running maximum of the account value. Inspired by \citet{milevsky} we consider a product under which the policyholder can withdraw a guaranteed amount set as a fraction $\gamma$ of the running maximum of the investment account. At maturity the remaining value is converted into a life-time annuity with a guaranteed consumption rate $L$. Such product could allow for higher consumption in the times of booming financial markets before locking the accumulated money into the fixed life-time annuity. This could be an example of an income drawdown scheme in retirement planning, see  \citet{emms}. We have to find the investment strategy $\pi$ under which the dynamics of the investment account fulfills
\beq\label{supwithdrawal0}
dX(t)&=&\pi(t)(\mu(t)dt+\sigma(t)dW(t))+(X(t)-\pi(t))r(t)dt\nonumber\\
&&-\gamma\sup_{s\in[0,t]}\{X(s)\}dt,\nonumber\\
\quad X(T)&=&La(T),
\eeq
where $a$ denotes the annuity factor
\beqo
a(T)=\be^\bq\Big[\int_T^\infty e^{-\int_T^sr(u)du}ds|\mathcal{F}_T\Big].
\eeqo
In the traditional approach in order to cover the minimum withdrawal rate a hedging fee is deducted from the account $X$. Our approach is to find a strategy $\pi$ under which $X$ is sufficient to cover the guaranteed stream of the cash flows.\\
\indent We can prove the following result.
\begin{prop}
Assume that (B1)-(B3) hold and that $\tilde{a}(T)$ is square integrable. The time-delayed BSDE
\beq\label{supwithdraw}
Y(t)&=&L\tilde{a}(T)+\int_t^T\gamma\sup_{u\in[0,s]}\big\{Y(u)e^{-\int_u^sr(v)dv}\big\}ds\nonumber\\
&&-\int_t^TZ(s)dW^\bq(s),\quad 0\leq t\leq T.
\eeq
has a unique square integrable solution $(Y,Z)$ for a sufficiently small $\gamma$ or $T$ under the requirement that $Y(0)\geq 0$. The solution $Y$ is strictly positive.
\end{prop}
\Proof The existence, uniqueness and integrability follow from Theorem 2.1. One can see easily that the pair $Y=Z=0$ does not satisfy the time-delayed BSDE \eqref{supwithdraw}. Strict positivity is the result of positivity of the terminal condition and the generator as we have
\beqo
Y(t)=\be^\bq\Big[L\tilde{a}(T)+\int_t^T\gamma\sup_{u\in[0,s]}\big\{Y(u)e^{-\int_u^sr(v)dv}\big\}ds|\bfi_t\Big]> 0.
\eeqo
\cbdu
We remark that for such a retirement product $\gamma$ is usually small, see \citet{milevskypos} and \citet{milevskyva}.\\
\indent Unfortunately, we cannot solve the equation \eqref{supwithdraw} explicitly. At the moment no algorithm exists for solving time-delayed BSDEs numerically. One could try following the scheme from \citet{bender} which is based on Picard iterations. A unique solution to the time-delayed BSDE \eqref{supwithdraw} could be obtained as the limit of the sequence of the processes satisfying
\beqo
Y^n(t)=\be^\bq\Big[L\tilde{a}(T)+\int_t^T\gamma \sup_{u\in[0,s]}\big\{Y^{n-1}(u)e^{-\int_u^sr(v)dv}\big\}ds|\bfi_t\Big],
\eeqo
and by approximating the expectation by some estimator based on Monte-Carlo simulations. The first hint for estimating the expected value could be to approximate the Brownian motion by a symmetric random walk as in \cite{maprotter}. The evaluations of the expectations conditioned on the whole past trajectory would be easy but computationally intensive. Such an algorithm would be feasible for a small time interval but impractical for longer durations due to an enormous number of trajectories that has to be generated. Our first simulation results, which we do not present here, show a good performance of such "naive" algorithm on a small time interval. If $r$ is a Markov process then one could try using the Markovian structure of the fully coupled forward-backward stochastic differential equation
\beqo
dY^n(t)&=&-\gamma Q^{n-1}(t)dt+Z^n(t)dW^\bq(t),\quad Y^n(T)=L\tilde{a}(r(T)),\\
Q^{n-1}(t)&=&\sup_{u\in[0,s]}\big\{Y^{n-1}(u)e^{-\int_u^sr(v)dv}\big\},
\eeqo
and represent $Y^n(t)=f(t,r(t),Q^{n-1}(t))$. The second hint could be to estimate $f$ by a least square Monte-Carlo, see \citet{bender}. However, our forward-backward structure is beyond the theory studied in the framework of FBSDE. The application of such algorithm would be far from obvious. Important questions of convergence still remain.\\
\indent Moving a step further, we could also try to find the investment strategy $\pi$ under which the dynamics of the investment account fulfills
\beq
dX(t)&=&\pi(t)(\mu(t)dt+\sigma(t)dW(t))+(X(t)-\pi(t))r(t)dt\nonumber\\
&&-\gamma\sup_{s\in[0,t]}\{X(s)\}dt,\nonumber\\
\quad X(T)&=&\gamma\sup_{0\leq s\leq T}\{X(s)\}a(T).
\eeq
Compared to \eqref{supwithdrawal0} the last withdrawal rate is now locked in the life-time annuity, see \citet{milevsky}. Things get much more complicated as the unique solution derived under Theorem 2.1 is $X=\pi=0$ and it is not clear at the moment how one should derive a non-zero solution (if such exists at all). More work on time-delayed BSDEs is needed in the future.

\section{Conclusion}
In this paper we have investigated novel applications of a new class of equations which we called time-delayed backward stochastic differential equations. We have discussed that a time-delayed BSDE may arise in finance when we want to find an investment strategy and an investment portfolio which should replicate a liability or meet a target depending on the applied strategy or the past values of the portfolio. The prime examples include a construction of a capital-protected investments and the hedging of a performance-linked pay-off. We have commented on participating contracts and variable annuities. We have pointed out an important area of possible applications of time-delayed BSDEs which has not been exploited in the literature so far and seems to be promising as far as further research is concerned. We would like to refer the reader to \citet{delong4} where applications of time-delayed BSDEs to dynamic pricing and recursive utilities are considered.

\end{document}